\documentclass[preprint,showpacs,preprintnumbers,amsmath,amssymb]{revtex4}

\usepackage{graphicx}
\usepackage{dcolumn}
\usepackage{bm}

\begin{document}

\preprint{}

\title{Hilbert-Schmidt  Orthogonality of $\det(\rho)$ and 
$\det(\rho^{PT})$ over the Two-Rebit Systems $\rho$ and Further 
Determinantal Moment Analyses}

\author{Paul B. Slater}%
\email{slater@kitp.ucsb.edu}
\affiliation{%
ISBER, University of California, Santa Barbara, CA 93106\\
}%
\date{\today}

\begin{abstract}
A complete description of the multitudinous ways in which quantum particles can be entangled requires the use of high-dimensional abstract mathematical spaces. We report here a particularly interesting feature of the nine-dimensional convex set--endowed with Hilbert-Schmidt (Euclidean/flat) 
measure--composed of two-rebit ($4 \times 4$) density matrices ($\rho$).
To each $\rho$ is assigned the product of its (nonnegative)
determinant $|\rho|$ and the determinant of its partial transpose 
$|\rho^{PT}|$--negative 
values of which, by the results of Peres and Horodecki, signify the entanglement of $\rho$. Integrating this product,  
$|\rho| |\rho^{PT}| =|\rho \rho^{PT}|$, over the nine-dimensional space,  using the indicated (HS) measure, we obtain the result zero. The two determinants, thus, form a pair of 
multivariate orthogonal polynomials with respect
to HS measure.  However, orthogonality does not hold, we find, if the symmetry of the nine-dimensional  two-rebit scenario is broken slightly, nor with the use of non-flat measures, such as the prominent Bures (minimal monotone)
measure--nor 
in the full HS extension to the 15-dimensional convex set of
two-qubit density matrices. We discuss relations--involving the HS moments of 
$|\rho^{PT}|$--to the long-standing problem of determining the probability that a generic pair of rebits/qubits is separable.
\end{abstract}


\maketitle
\section{Introduction and Statement of Main Results}
The simplest form of finite-dimensional system 
capable of exhibiting the fundamental, holistic 
property of entanglement--"that feature of quantum formalism which makes it impossible to simulate quantum correlations within any classical formalism" \cite{clan}--is that composed of a pair of two-level
systems (quantum bits or "qubits") \cite{ingemarkarol}. The joint state of two qubits is describable  ($2 \times 2 = 4$) 
by a $4 \times 4$ 
density matrix $\rho$, that is 
a Hermitian matrix (its transpose equalling its complex conjugate) 
having its four nonnegative diagonal entries (probabilities)--as well as its four nonnegative eigenvalues--summing to one \cite{augusiak}. 
The entirety of such $4 \times 4$ density matrices with their entries restricted to real values--which will be our principal subject of analysis here--forms a  nine-dimensional convex set. We note that $2 \times 2$ density matrices with real entries have been termed "rebits" \cite{carl}.

Endowing the nine-dimensional set of pairs of rebits with Hilbert-Schmidt (HS) (Euclidean/flat) measure \cite[sec. 14.3]{ingemarkarol}  \cite{szHS,andai}, we 
arrive at  the following trio of computational results \cite{zeromean}.
The three are: (1) the average value or mean (first moment) of the (necessarily nonnegative) 
determinant of $\rho$, denoted $|\rho|$ ($\equiv \det(\rho)$), 
is $\frac{1}{2288}= (2^4 \cdot 11 \cdot 13)^{-1}$; (2) the mean of the determinant $|\rho^{PT}|$ 
of the partial transpose (PT) of $\rho$--negative values of which, by the celebrated results of Peres \cite{asher} and the Horodecki family 
\cite{michal}, are fully equivalent (for qubit-qubit systems--as well  as
qubit-{\it qutrit} 
systems, representable by $6 \times 6$ density matrices)
to the entanglement/nonseparability of $\rho$--is $-\frac{1}{858}= -(2 \cdot 3 \cdot 11 \cdot 13)^{-1}$; and (3) the mean of 
the product of these two determinants, that is $|\rho| |\rho^{PT}|$ ($ =|\rho \rho^{PT}|$ by the Cauchy-Binet [Gram's] Theorem), is zero. 
(The partial transpose--identifiable with the antiunitary operation of time-reversal \cite{bruss} (cf. \cite{avron1,avron2})--of a $4 \times 4$ matrix can be obtained simply by transposing in place its four $2 \times 2$ blocks.)

So, by assigning
to each generic real $4 \times 4$ 
density matrix $\rho$, a value equal to this product of determinants, we find that the
total (but yet undetermined, necessarily negative) 
value allotted to the entangled states exactly cancels the (necessarily nonnegative) value allotted 
to the disentangled/separable states. This result appears to be special to the 
HS measure, as the analogous trio of results using the well-known
Bures (minimal monotone) measure 
\cite[sec. 14.4]{ingemarkarol} \cite{szBures} is $\frac{1}{8192}=2^{-13}$ and, numerically, 
-0.0030959720 and $-1.124478 \cdot 10^{-7}$ \cite{zeromean}.
It is a natural hypothesis--but one apparently presenting substantial computational challenges to test--that this zero-mean (HS) result extends 
to the standard 
15-dimensional convex set of  $4 \times 4$ density matrices with
arbitrary {\it complex} off-diagonal entries \cite[sec. 2]{quartic}. (From a "physical point of view one should in general consider the entire set of complex density matrices" \cite{szHS}. We can not, however, unqualifiedly, 
straightforwardly extend the interpretation of the zero-product-mean hypothesis to bipartite quantum systems describable by density matrices of dimensions greater than four. Then, $\rho^{PT}$ can have more than one negative eigenvalue, with an even number of such eigenvalues yielding a positive value for  the 
determinant $|\rho^{PT}|$ (cf. \cite{augusiak}). Thus, a nonnegative value of $|\rho^{PT}|$ would not necessarily indicate that $\rho$ is separable.)

The zero-product-mean HS result does not hold, in general, however, 
for arbitrary two-qubit scenarios. 
This can be seen by slightly modifying in each of three ways, our 
basic nine-dimensional two-rebit 
example--thereby diminishing its inherent symmetry.
Firstly, if we set one off-diagonal pair of entries ($\rho_{34}=\rho_{43}$, say) of $\rho$ to zero, giving us an 8-dimensional convex set of density matrices, we obtain a mean of the probability distribution over $|\rho|$ of $\frac{1}{4752}= (2^4 \cdot 3^3 \cdot 11)^{-1} 
\approx 0.000210438$, for 
$|\rho^{PT}|$, a mean of $-\frac{13}{9504} = -13 (2^5 \cdot 3^3 \cdot 11)^{-1} \approx -0.00136785$, and for the product
of the two determinants, a {\it nonzero} mean of $\frac{1}{2196480}= (2^{10} \cdot 3 \cdot 5 \cdot 11 \cdot 13)^{-1} \approx 4.55274 \cdot 10^{-7}$.  Secondly, if we
{\it increase} the dimensionality from nine to ten by letting $\rho_{34}$ 
and $\rho_{43}$ be arbitrary complex conjugates of each other,
then, the trio of means, numerically, is 0.000412154, -0.00082468, and 
(small, but apparently nonzero) $3.6035 \cdot 10^{-8}$.
Thirdly, for the eight-dimensional set of 
$4 \times  4$ density matrices with real entries consisting of those minimally degenerate matrices having one eigenvalue zero--which constitute  the boundary of the basic nine-dimensional set \cite{sbz}--the mean of the product  of the three nonzero eigenvalues is $\frac{1}{66} \approx 0.0151515$, the mean of  
$|\rho^{PT}|$  is 
$-\frac{5}{2376} \approx -0.00210438$, while the mean of the product of 
$|\rho^{PT}|$ and 
the three nonzero eigenvalues  is also nonzero, that is, 
$-\frac{1}{47520} \approx -0.0000210438$. (Of course, in this last 
minimally degenerate 
analysis, the determinant $|\rho|$, equalling the product of its
{\it four} eigenvalues, is zero. So, we replace it, for our purposes, by the product of the three generically nonzero eigenvalues.) We, additionally, have found that {\it no} pair of principal ($4 \times 4$, $3 \times 3$, $2 \times 2$) minors, other than $|\rho|$ and $|\rho^{PT}|$--one minor of $\rho$ and the other of $\rho^{PT}$--are orthogonal over the nine-dimensional set with
respect to HS measure.

Since both $|\rho|$ and $|\rho^{PT}|$ in the generic (two-rebit) nine-dimensional case  are polynomials in nine variables, it is suggestive, at least, to consider these
two determinants as a pair of multivariate orthogonal polynomials (MOPS)\cite{dumitriu,dunkl2,griffithsspano} with respect to the HS measure. 
Griffiths and Span{\`o} have reviewed multivariate orthogonal polynomials with respect to weight measures given by Dirichlet distributions over simplices (such as that to be  given below by (\ref{andaiDirichlet}) \cite{andai} over the 3-dimensional simplex formed by the diagonal entries of $4 \times 4$ density matrices \cite{griffithsspano}).

Let us also note, in our basic nine-dimensional scenario, that the expected value with respect  to the Hilbert-Schmidt measure of the determinant of the  {\it commutator} (appearing in the von Neumann equation for the 
time evolution of $\rho$), 
$[\rho,\rho^{PT}] = \rho \rho^{PT}- \rho^{PT} \rho$, is zero too.
(Jarlskog found the the determinant of the commutator of mass matrices  vanishes "if and only if there is no 
{\it CP} nonconservation" \cite{jarlskog}.)
The analogous result was nonzero, that is $\frac{79}{27675648} \approx 2.8545 \cdot 10^{-6}$, when, 
once again, we broke the symmetry by setting $\rho_{34}=\rho_{43}=0$.

The range of possible values of 
$|\rho|$ is $[0,2^{-8}]$, that of 
$|\rho^{PT}|$ is $[-2^{-4},2^{-8}]$, while that of the product  
$|\rho| |\rho^{PT}|$, in the nine-dimensional scenario, we find, is $[-2^{-12} \cdot 3^{-3},2^{-16}]$.
This last upper limit is reached, clearly, by the fully mixed 
(classical) state--the diagonal density matrix with diagonal entries 
(probabilities) all equal to 
$\frac{1}{4}$. An instance of a density matrix attaining the last lower limit is 
\begin{equation} \label{NewMatrix}
\rho= 
\left(
\begin{array}{cccc}
 \frac{1}{6} & -\frac{1}{6 \sqrt{2}} & \frac{1}{6
   \sqrt{2}} & \frac{1}{12} \left(-1+\sqrt{3}\right) \\
 -\frac{1}{6 \sqrt{2}} & \frac{1}{3} & \frac{1}{12}
   \left(-1-\sqrt{3}\right) & -\frac{1}{6 \sqrt{2}} \\
 \frac{1}{6 \sqrt{2}} & \frac{1}{12}
   \left(-1-\sqrt{3}\right) & \frac{1}{3} & \frac{1}{6
   \sqrt{2}} \\
 \frac{1}{12} \left(-1+\sqrt{3}\right) & -\frac{1}{6
   \sqrt{2}} & \frac{1}{6 \sqrt{2}} & \frac{1}{6}
\end{array}
\right).
\end{equation}
Now, the determinant $|\rho|$  is 
$\frac{1}{576} \left(2 \sqrt{3}-3\right) \approx 0.000805732$
and
$|\rho^{PT}| =\frac{1}{576} \left(-3-2 \sqrt{3}\right) \approx -0.0112224$ 
(their product being the indicated lower limit 
$-\frac{1}{110592} \approx -9.04225 \cdot 10^{-6}$).
Both $\rho$  and $\rho^{PT}$ have three identical eigenvalues 
($\frac{1}{12} \left(3-\sqrt{3}\right) \approx 0.105662$ for $\rho$, and 
$\frac{1}{12} \left(3+\sqrt{3}\right) \approx 0.394338$ for $\rho^{PT}$.
The isolated eigenvalue for $\rho$ is $\frac{1}{4} \left(1+\sqrt{3}\right) 
\approx 0.683013$, and for $\rho^{PT}$, $\frac{1}{4} \left(1-\sqrt{3}\right) \approx -0.183013$. The purity (index of coincidence \cite[p. 56]{ingemarkarol}), $\mbox{Tr}(\rho^2)$, of (\ref{NewMatrix}) equals $\frac{1}{2}$, so its inverse (the participation 
ratio \cite{ingemarkarol,ZHSL}) is 2.
\subsection{Failure of Hilbert-Schmidt 
orthogonality in generic 15-dimensional two-qubit case}
With the considerable computational (Mathematica) assistance of Michael Trott, we investigated the orthogonality hypothesis in the more demanding 15-dimensional generic two-qubit scenario, utilizing the Euler-angle parameterization of Tilma, Byrd and Sudarshan
\cite{tbs},  together with the appropriate 
formulas of {\.Z}yczkowski and Sommers 
\cite[eq. (3.11)]{szHS}. For the expected value of 
$|\rho| |\rho^{PT}|$, we obtained the {\it nonzero} value of 
$-\frac{1}{4576264} \approx -2.18519 \cdot 10^{-7}$, and for the mean of 
$|\rho^{PT}|$, 
the more negative value, $-\frac{7}{3876} \approx -0.00180599$. (Normalizing the two quantitates $|\rho|$ and $|\rho^{PT}|$ to have unit length [the latter numerically], we find that their inner product is approximately -0.07. This translates into an angle of 94 degrees, close to orthogonality.) are, since it appears to be even more computationally challenging to normalize  
$|\rho^{PT}|$ so that its average HS length is unity.) (However, we are not presently able to evaluate how close to
orthogonality the determinants $|\rho|$ and $|\rho^{PT}|$ are, since it appears to be even more computationally challenging to normalize  
$|\rho^{PT}|$ so that its average HS length is unity.) 
The mean of $|\rho|$ is known \cite[eq. (3.11)]{szHS} to be
$\frac{1}{3876} = (2^2 \cdot 3 \cdot 17 \cdot 19)^{-1}  \approx  0.000257998$.

\subsection{Separability probabilities}
It has been a relatively long-standing problem of "philosophical, practical and physical" significance \cite{ZHSL} (in the recently-burgeoning  field of quantum information \cite{mikeike}) 
to determine the probability (with respect to a number of measures, such as the HS and Bures) that a pair of qubits is separable \cite{ZHSL,advances,slater833,slaterJGP2}. 
In this regard, it would be of interest to, in some manner, combine (to achieve a fuller understanding of the "geometry of quantum states" \cite{ingemarkarol}) 
the trio of HS (first) moment (mean) results given above with two  theorems (apparently the only ones yet developed) pertaining to 
entanglement in terms of the Hilbert-Schmidt measure (cf. \cite{slaterJGP2}). 
\subsubsection{Existing theorems}
One of these theorems states that the probability that a generic 
bipartite state (of arbitrary dimension) has a positive partial transpose (PPT) is {\it twice} the probability
that a generic boundary (minimally degenerate [one eigenvalue zero]) 
state has a PPT. (For the proof, it was established that the convex set of mixed PPT states is "pyramid-decomposable and hence is a body of constant height" \cite{sbz}.) The other (but now dimension-specific) 
theorem--derived by enforcing the nonnegativity of pairs of principal $3 \times 3$ minors of $\rho^{PT}$--is that
the probability that a generic real $4 \times 4$ density matrix has 
a PPT (or, equivalently, by the Peres-Horodecki results, is separable)  
is no greater than $\frac{1129}{2100} \approx 0.537619$, nor no less (using the concept of {\it absolute} separability \cite{kus} and the important Verstraete-Audenaert-de Moor bound \cite{ver}) than
$\frac{6928 -2205 \pi}{2^{9/2}} \approx 0.0348338$ \cite{advances}. 
(An absolutely separable state is one that can not be entangled by any unitary transformation.) From these two theorems together, one easily obtains the corollary that the HS separability probability of a generic  two-rebit minimally degenerate state is no greater than 
$\frac{1129}{4200} \approx  0.26881$. 
\subsection{Higher-order moments}
In addition to the trio of HS means ({\it first} moments) given above, we have been able to compute exactly several higher-order moments, as well, for the nine-dimensional scenario. The corresponding trio of {\it second} (raw/non-central) moments is
$\frac{1}{2489344} 
= (2^{10} \cdot 11 \cdot 13 \cdot 17)^{-1} \approx 4.01712 \cdot 10^{-7}$, $\frac{27}{2489344} =\frac{3^3}{2^{10} \cdot 11 \cdot 13 \cdot 17}\approx 0.0000108462$, and $\frac{7}{5696343244800}=\frac{7}{2^{18} \cdot 3^2  \cdot 5^2 \cdot 13 \cdot 17 \cdot 19 \cdot 23} \approx 1.2288585 \cdot 10^{-12}$. From our several results, we can deduce that the 
generic two-rebit Hilbert-Schmidt correlation  between $|\rho|$ and $|\rho^{PT}|$ is positive, that is 
$\frac{544}{5 \sqrt{91191}} \approx 0.360291$. 

Further, making use of the generalized normalization constants 
\cite[eq. (4.3)]{szHS}, we obtain that the $m$-th moment of the probability distribution over $|\rho|$ is 
\begin{equation} \label{generalformula}
\zeta_{m/real}^{'}=945 \frac{4^{3-2 m} \Gamma (2 m+2) \Gamma (2 m+4)}{\Gamma (4 m+10)}.
\end{equation}
Let us also note that the counterpart of this result 
for the 15-dimensional generic complex two-qubit states is 
\begin{equation}
\zeta_{m/complex}^{'}=108972864000 \frac{ \Gamma (m+1) \Gamma (m+2) \Gamma (m+3) \Gamma
   (m+4)}{\Gamma (4 (m+4))}.
\end{equation}
\subsubsection{Application of Chebyshev inequality}
The skewness ($\gamma_1$) of the Hilbert-Schmidt probability distribution over $|\rho^{PT}|$ is negative (as well as all odd moments [$m=1,3,5,7,9$] so far computed), that is, -3.13228--so, the left tail of the distribution is more pronounced than the right tail--while its kurtosis ($\gamma_2$), a measure of "peakedness" is quite high, 17.6316.
(Higher kurtosis indicates that more of the variance is the result of infrequent extreme deviations than frequent modestly sized 
deviations.)
From the first two moments, we obtain the variance
\begin{equation}
\sigma^2=\frac{30397}{3203785728} \approx 9.487838 \cdot 10^{-6}.
\end{equation}
Application of the standard-form one-sided Chebyshev inequality \cite{marshall} 
(we perform a linear transformation,  so that negative values of
$|\rho^{PT}|$ are mapped to [0,1]), then, yields an upper bound on the Hilbert-Schmidt separability probability of the  two-rebit density matrices of 
$\frac{30397}{34749} \approx 0.874759$. This is a substantially weaker upper bound, however, than that of $\frac{1129}{2100} \approx 0.537619$ established in \cite{advances}, by enforcing the nonnegativity of pairs of $3 \times 3$ principal minors of the partial transpose, and even weaker than 
$\frac{1024}{135 \pi^2} \approx 0.76854$, obtained by requiring the nonnegativity
of all six $2 \times 2$ principal minors of the partial transpose 
\cite{advances}.
\subsection{Fits of probability distributions to computed moments}
\subsubsection{Beta distribution fit to first two moments} \label{BetaSection}
\begin{figure}
\includegraphics[scale=2]{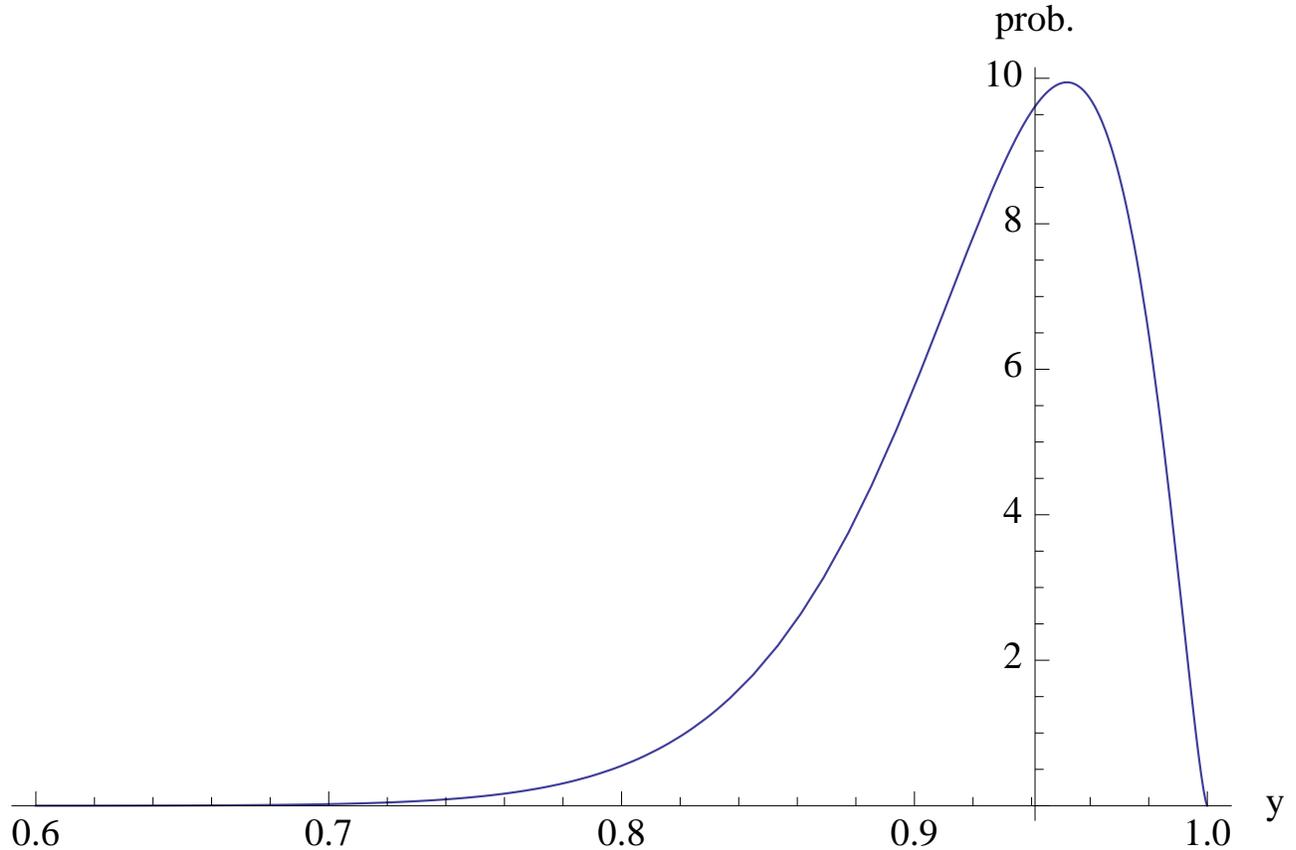}
\caption{\label{fig:goodbetafit}The (two-parameter) beta probability distribution (\ref{goodbetafit2}) that closely (within $98.6\%$) reproduces the (linearly transformed to the interval [0,1]) first exact nine moments of 
$|\rho^{PT}|$. The vertical axis corresponds to the separable-entangled boundary ($y=16/17 \approx 0.941176$), so the mode of the distribution (located at $y=0.95206948$, corresponding to 
$|\rho^{PT}| = 0.000723364$) lies within the separable region.}
\end{figure}

\begin{figure}
\includegraphics[scale=2]{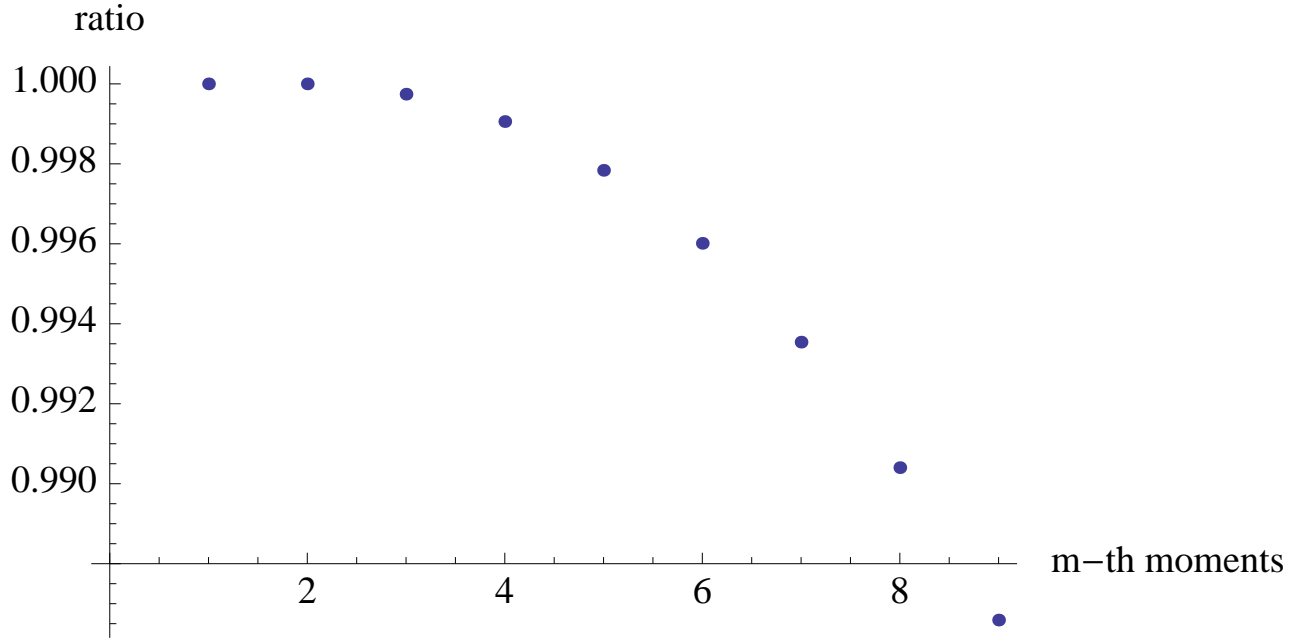}
\caption{\label{fig:ratiosofmoments}Ratios of the $m$-th exact moment of $|\rho^{PT}|$ (linearly transformed to the interval [0,1]) to the $m$-th moment of the beta probability distribution (\ref{goodbetafit2}) fitted to the first and second moments}
\end{figure}

Further, we linearly mapped $|\rho^{PT}| \in [-\frac{1}{16},\frac{1}{256}]$ to $y$, so that $y \in  [0,1]$, and transformed its exact first nine moments  accordingly. 
Then we exactly fit the first and second such moments to a basic (two-parameter)
beta distribution, giving us a probability distribution of the form 
(Fig.~\ref{fig:goodbetafit}),
\begin{equation} \label{goodbetafit2}
P_{|\rho^{PT}|}(y) = \frac{y^{(a-1)} (1-y)^{(b-1)}}{B(a,b)};\hspace{.3in} a= \frac{15171156}{516749} \approx 29.3588, 
b= \frac{5018013}{2066996} \approx 2.42768, 
\end{equation}
where $B(a,b)$ is the beta function.
The ratios of the next six transformed moments ($m=3,\ldots,8$) to the corresponding moments of 
this beta distribution all rather remarkably lie between 0.99 and 1 (with the ratio of the ninth moments
being 0.986) (Fig.~\ref{fig:ratiosofmoments}). Integrating the distribution (\ref{goodbetafit2}) over the interval
$y \in [\frac{16}{17},1]$, we obtain an associated separability probability estimate
of 0.4183149.

We did similarly linearly transform the product of $|\rho|$ and $|\rho^{PT}|$ to lie in the unit interval [0,1], and exactly fit to the so-transformed first two moments, a beta distribution of the 
form (cf. (\ref{goodbetafit2}))
\begin{equation} \label{goodbetafit4}
P_{|\rho| |\rho^{PT}|}(y) = \frac{y^{(a-1)} (1-y)^{(b-1)}}{B(a,b)};\hspace{.3in} a= \frac{2392921}{57792} \approx 41.4057, 
b= \frac{21536289}{308224} \approx 68.8722.
\end{equation}
The associated probability (over the separability domain 
$[\frac{16}{43},1]$) was computed as 0.49331935.
Using numerical methods, we found that the (transformed) third 
moment of the distribution of the product $|\rho| |\rho^{PT}|$
was approximately $99.84\%$ as large as the third moment of (\ref{goodbetafit4}),
but the fourth moment was only $62.25\%$ as large. So, this fit to a beta distribution certainly appears to be inferior to the earlier one 
(\ref{goodbetafit2}) (Figs.~~\ref{fig:goodbetafit} and \ref{fig:ratiosofmoments}) 
based on fitting the first two moments of $|\rho^{PT}|$. 
\subsubsection{Use of two general moment-fitting procedures}
\begin{figure}
\includegraphics[scale=2]{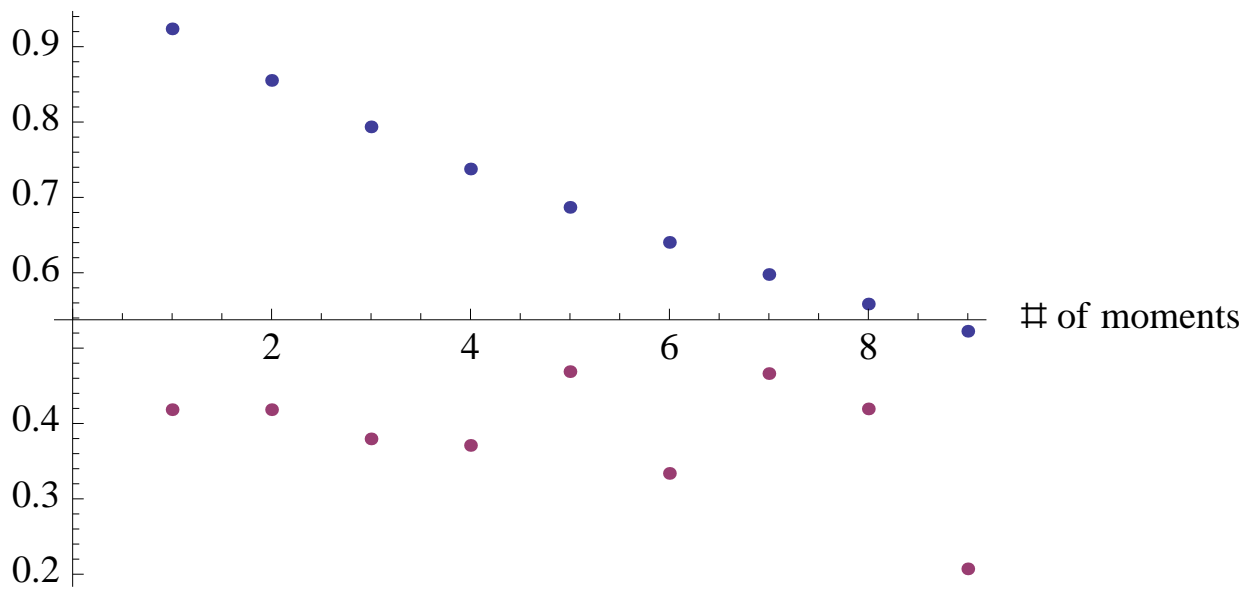}
\caption{\label{fig:CDFHP}Separability probability estimates for the generic real $4 \times 4$ density matrices based on
differing numbers ($m=1,\ldots,9$) 
of exact moments of $|\rho^{PT}|$, using the 
nonparametric reconstruction procedures of Mnatsakanov \cite{mnatsakanov2}--giving the higher (blue) set of nine points--and the  polynomial adjustment of the baseline density 
methodology  of Ha and Provost \cite{HP}. The horizontal axis is drawn to intercept the vertical at $\frac{1129}{2100} \approx 0.537619$, the least upper bound so far established \cite{advances}.}
\end{figure}

One can use the calculated exact nine moments of $|\rho^{PT}|$ in certain formulas for cumulative distribution functions, in order to approximate the desired, specific (separability) probability
that $|\rho^{PT}|$ is greater than zero \cite[eq. (2)]{mnatsakanov2,HP} 
(cf. \cite{bertsimas}). 
In Fig.~~\ref{fig:CDFHP} we display together one such set of  nine (greater) separability probability estimates based on the nonparametric (stable approximant)
reconstruction approach of Mnatsakanov \cite{mnatsakanov2}, 
along with the analogous first nine (lesser) estimates
using the orthogonal-polynomial-based methodology of Provost and Ha \cite[eq. (3.5)]{HP}. (In the latter approach, we used the well-fitting beta distribution (\ref{goodbetafit2}) as a "baseline" density that is 
adjusted by associated modified Jacobi orthogonal polynomials.) 
It is clear that additional (exact) moments ($m>9$) are certainly needed to more satisfactorily
sharpen the generic  two-rebit separability probability estimates in Fig.~~\ref{fig:CDFHP}. Then, one might be able to obtain convincing evidence--in the absence of the desired, but highly challenging exact symbolic computation--for the true (hypothetically exact 
\cite{advances,slater833,slaterJGP2}) value of the separability probability. (Perhaps, somewhat relatedly, Gurvits has shown "that the weak membership problem for the convex set of separable normalized bipartite density matrices is NP-hard'' \cite{Gurvits}.)

\section{Methodologies Employed} \label{methodology}
\subsection{Density Matrix Parameterizations}
Our analysis proceeds in the framework of the
Bloore (or correlation coefficient) parameterization \cite{bloore,joe,slaterPRA2} of the $4 \times 4$ density matrices ($\rho$)  which allows us (in the generic two-rebit case of immediate interest here) to work primarily in seven dimensions, rather than  the nine naively expected. Also, in our computations, we still 
further reparametrize three ($z_{13}, z_{14}, z_{24}$)  of the six correlations
\begin{equation}
z_{ij}=\frac{\rho_{ij}}{\sqrt{\rho_{ii} \rho_{jj}}}, \hspace{.1in} 1 \leq i <j \leq 4, \hspace{.2in} 
z_{ij} \in [-1,1]
\end{equation}
in terms of {\it partial} correlations \cite{joe}, allowing certain requisite integrations to be performed simply over six-dimensional hypercubes, rather than more complicated domains \cite{advances}. (Alternatively, and reasonably computationally competitively, one may utilize the cylindrical algebraic decomposition \cite{cylindrical}
to define the integration limits [as indicated in 
\cite[sec. II]{slaterPRA2}] that specify the domain of feasible density matrices, directly within the Bloore-type framework, without  transforming to partial correlations.) 

The computation of the $m$-th Hilbert-Schmidt moment over $|\rho^{PT}|$ is carried out in two stages. (The moments of the distribution 
of purity $P(\rho)= \mbox{Tr} \rho^2$ for quantum states with respect to the {\it Bures} measure have recently been determined by Osipov, Sommers and {\.Z}yczkowski  \cite{osipov}, while Giraud has investigated  the Hilbert-Schmidt counterparts \cite{giraud1,giraud2}.) In the first stage,
we perform an integration over the six-dimensional hypercube $[-1,1]^6$ of the $m$-th power of a (transformed) polynomial ($\tilde{P}$)--proportional to $|\rho^{PT}|$--in seven variables (\cite[eq. (7)]{slaterPRA2}). (The proportionality factor is $(\rho_{22} \rho_{33})^{2 m}$.) 
The free (unintegrated) variable is of the form
\begin{equation} \label{substitution}
\mu=\sqrt{\frac{\rho_{11} \rho_{44}}{\rho_{22} \rho_{33}}},
\end{equation}
where the $\rho_{ii}$'s are the diagonal entries of $\rho$. (In the related study \cite{slaterPRA2}, $\nu=\mu^2$ was used as the principal 
variable, and in 
\cite{advances}, $\xi=\log{\mu}$.) 
We have that (before the transformation (\ref{transformationJoe}) to partial correlations, yielding $\tilde{P}$) the multivariate polynomial 
(proportional to $|\rho^{PT}|$)
\begin{equation} \label{PT}
P=-z_{{14}}^2 \mu ^4+2 z_{{14}} \left(z_{{12}}
   z_{{13}}+z_{{24}} z_{{34}}\right) \mu ^3 + \left(V +W \right) \mu ^2 + 2
   z_{{23}} \left(z_{{12}} z_{{24}}+z_{{13}}
   z_{{34}}\right) \mu -z_{{23}}^2 ,
\end{equation}
where
\begin{displaymath}
V=\left(z_{{34}}^2-1\right) z_{{12}}^2-2
   \left(z_{{14}} z_{{23}}+z_{{13}} z_{{24}}\right)
   z_{{34}} z_{{12}}+z_{{14}}^2
   z_{{23}}^2-z_{{24}}^2-z_{{34}}^2
\end{displaymath}
and 
\begin{displaymath}
W=-2 z_{{13}}
   z_{{14}} z_{{23}} z_{{24}}+z_{{13}}^2
   \left(z_{{24}}^2-1\right)+1.
\end{displaymath}
The transformation of the three correlations $z_{13}, z_{14}, z_{24}$ to partial correlations (denoted $z_{13,2},z_{24,3},z_{14,23}$)
takes the form \cite{joe}
\begin{equation} \label{transformationJoe}
z_{{14}}\to z_{{12}} z_{{23}}
   z_{{34}}+\sqrt{z_{{12}}^2-1} \sqrt{z_{{23}}^2-1}
   z_{{13},2} z_{{34}}+z_{{12}} \sqrt{z_{{23}}^2-1}
   \sqrt{z_{{34}}^2-1} z_{{24},3} +
\end{equation}
\begin{displaymath}
\sqrt{z_{{12}}^2-1} \sqrt{z_{{34}}^2-1}
   \sqrt{z_{{13},2}^2-1} \sqrt{z_{{24},3}^2-1}
   z_{14,23}+\sqrt{z_{{12}}^2-1} z_{{23}}
   \sqrt{z_{{34}}^2-1} z_{{13},2} z_{{24},3},
\end{displaymath}
\begin{displaymath}
z_{{13}}\to z_{{12}} z_{{23}}+\sqrt{z_{{12}}^2-1}
   \sqrt{z_{{23}}^2-1} z_{{13},2},z_{{24}}\to z_{{23}} z_{{34}}+\sqrt{z_{{23}}^2-1}
   \sqrt{z_{{34}}^2-1} z_{{24},3}.
\end{displaymath}
The jacobian for this transformation is (note that one of the six 
variables [or three partial correlations]--$z_{14,23}$--is absent)
\begin{equation}
J(z_{12},z_{23},z_{34},z_{13,2},z_{24,3}) =\left(z_{{12}}^2-1\right) \left(z_{{23}}^2-1\right)
   \left(z_{{34}}^2-1\right) \sqrt{z_{{13},2}^2-1}
   \sqrt{z_{{24},3}^2-1}.
\end{equation}
\subsection{Intermediate functions/polynomials and their coefficients} \label{InterSection}
For the $m$-th moment ($Moment_{m} \equiv \zeta_m^{'}$), 
the indicated six-dimensional integration 
of $P^{m}$ in now reparameterized (partial correlation) form 
$\tilde{P}^{m}$ over the hypercube defined by $z_{12}, z_{23}, z_{34}, z_{13,2}, z_{24,3}, z_{14,23} \in [-1,1]$ takes the form--including a normalization factor of $\frac{27}{32 \pi^2}$--the ("intermediate function") result 
\begin{equation}
I_{m}(\mu)= \frac{27}{32 \pi^2} \int_{-1}^{1} \int_{-1}^{1} \int_{-1}^{1} \int_{-1}^{1} \int_{-1}^{1} \int_{-1}^{1} 
\end{equation}
\begin{displaymath}
J(z_{12},z_{23},z_{34},z_{13,2},z_{24,3}) [\tilde{P}(z_{12},z_{23},z_{34},z_{13,2},z_{24,3}, z_{14,23})]^m \mbox{d} z_{12} \mbox{d}z_{23}  \mbox{d} z_{34} \mbox{d} z_{13,2}
\mbox{d} z_{24,3}  \mbox{d}  z_{14,23}.
\end{displaymath}

\begin{figure}
\includegraphics[scale=2]{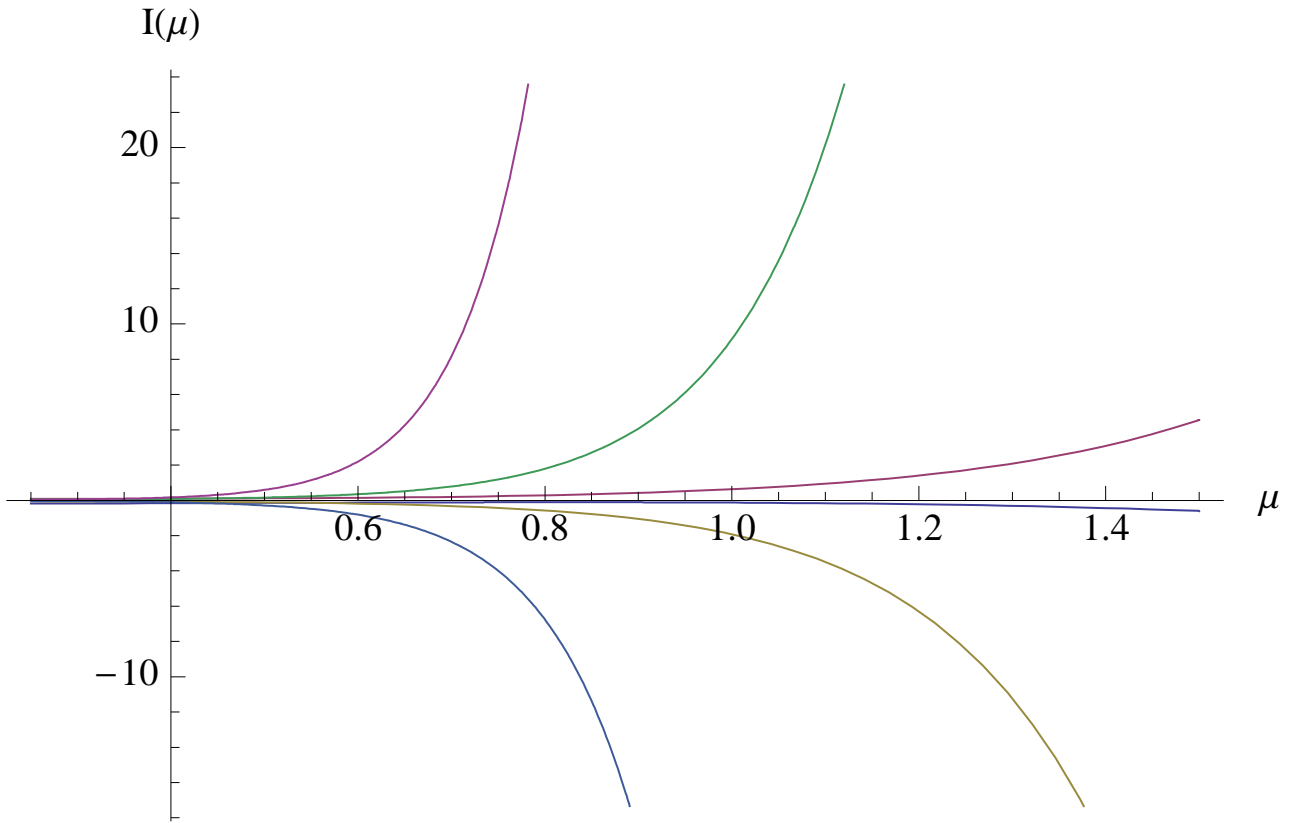}
\caption{\label{fig:intermediate}The six intermediate functions $I_{m}(\mu), m=1,...,6$. The curves for even $m$ curve upward, for odd $m$ downward, with the steepness of the curves increasing with $m$.}
\end{figure}

For the first ($m=1$) moment, we have the result 
(Fig.~\ref{fig:intermediate})
\begin{equation} \label{inter1}
I_{1}(\mu)=-\frac{\mu ^4}{5}+\frac{34 \mu ^2}{125}-\frac{1}{5},
\end{equation}
for the second moment ($m=2$),
\begin{equation} \label{inter2}
I_{2}(\mu)=\frac{3 \mu ^8}{35}-\frac{12 \mu ^6}{875}+\frac{20898 \mu
   ^4}{42875}-\frac{12 \mu ^2}{875}+\frac{3}{35},
\end{equation}
and for the third ($m=3$),
\begin{equation} \label{inter3}
I_{3}(\mu)=-\frac{\mu ^{12}}{21}-\frac{54 \mu ^{10}}{875}-\frac{27873 \mu
   ^8}{42875}-\frac{466876 \mu ^6}{1157625}-\frac{27873 \mu
   ^4}{42875}-\frac{54 \mu ^2}{875}-\frac{1}{21}.
\end{equation}
At this point, we omit terms of lower order $2 j$ than $2 m$, since their coefficients--in a symmetrical manner--match the coefficients $C_{4m-2j}(m)$.
Then, 
\begin{equation}
I_{4}(\mu)=\frac{\mu ^{16}}{33}+\frac{584 \mu ^{14}}{5775}+\frac{278884 \mu
   ^{12}}{282975}+\frac{8984 \mu ^{10}}{4851}+\frac{65788454 \mu
   ^8}{20543985}+\dots,
\end{equation}
\begin{equation}
I_{5}(\mu)=-\frac{3 \mu ^{20}}{143}-\frac{18 \mu ^{18}}{143}-\frac{70881 \mu
   ^{16}}{49049}-\frac{2178728 \mu ^{14}}{441441}-\frac{59472398 \mu
   ^{12}}{4855851}-\frac{4103383444 \mu ^{10}}{273546273}+\ldots,
\end{equation}
\begin{equation}
I_{6}(\mu)=\frac{\mu ^{24}}{65}+\frac{2556 \mu ^{22}}{17875}+\frac{5454 \mu
   ^{20}}{2695}+\frac{3359372 \mu ^{18}}{315315} +
\end{equation}
\begin{displaymath}
+\frac{3273117 \mu
   ^{16}}{86515}+\frac{597414184 \mu ^{14}}{7872865}+\frac{173821048732
   \mu ^{12}}{1771394625} +\ldots,
\end{displaymath}
\begin{equation}
I_{7}(\mu)= -\frac{\mu ^{28}}{85}-\frac{4298 \mu ^{26}}{27625}-\frac{826637 \mu
   ^{24}}{303875}-\frac{165865636 \mu ^{22}}{8204625}-\frac{71226035 \mu
   ^{20}}{722007} -
\end{equation}
\begin{displaymath}
-\frac{1947049760374 \mu
   ^{18}}{6711055065}-\frac{93373201818911 \mu
   ^{16}}{167776376625}-\frac{33225665966177656 \mu
   ^{14}}{48487372844625} \ldots,
\end{displaymath}
\begin{equation}
I_{8}(\mu)= \frac{3 \mu ^{32}}{323}+\frac{6672 \mu ^{30}}{40375}+\frac{12986136 \mu
   ^{28}}{3674125}+\frac{4250871568 \mu
   ^{26}}{121246125}+\frac{3319251741068 \mu
   ^{24}}{14670781125}+
\end{equation}
\begin{displaymath}
 +\frac{755365923834768 \mu
   ^{22}}{826454003375}+\frac{2024301386770232 \mu
   ^{20}}{826454003375}+\frac{61510285844520752 \mu
   ^{18}}{14049718057375}+\frac{3853435310162220966 \mu
   ^{16}}{724564031244625} \ldots,
\end{displaymath}
and
\begin{equation}
I_{9}(\mu)= -\frac{\mu ^{36}}{133}-\frac{9774 \mu ^{34}}{56525}-\frac{651051 \mu
   ^{32}}{145775}-\frac{8355664 \mu ^{30}}{146965}-\frac{18384996780 \mu
   ^{28}}{39122083}-\frac{4848288282648 \mu
   ^{26}}{1944597655}-
\end{equation}
\begin{displaymath}
-\frac{133915228926036 \mu
   ^{24}}{15026436425}-\frac{61222919937476688 \mu
   ^{22}}{2809943611475}-\frac{396008663496240078 \mu
   ^{20}}{10677785723605}-\frac{2103161056387491292 \mu
   ^{18}}{47564681859695}  \ldots
\end{displaymath}
For the nine cases ($m=1,...,9$) we have so far been able--with considerable computational expense--to explicitly compute, the coefficients of the corresponding $4m$-degree even 
polynomials $I_{m}(\mu)$ are, as already indicated, symmetric--for reasons not immediately apparent--around 
the $\mu^{2 m}$ term. 
\subsubsection{Formulas for the coefficients of the intermediate functions and their root and pole structure}
The constant terms (which equal the 
coefficients of the $\mu^{4 m}$
term) of the intermediate functions, used in the computation of the moments of $|\rho^{PT}|$,  are expressible as
\begin{equation} \label{coefficient0}
C_{0}(m)=C_{4m}(m)=\frac{3 (-1)^m}{4 \left(m+\frac{1}{2}\right) \left(m+\frac{3}{2}\right)}.
\end{equation}
Additionally, the coefficients of the second and $(4 m-2)$ terms are
\begin{equation} \label{coefficient2}
C_{2}(m)=C_{4m-2}(m)=\frac{3 (-1)^m m (2 m (4 m-5)-15)}{100 \left(m-\frac{1}{2}\right)
   \left(m+\frac{1}{2}\right) \left(m+\frac{3}{2}\right)}.
\end{equation}
Further, the coefficients of the fourth and $(4 m-4)$ terms are
\begin{equation} \label{coefficient4}
C_{4}(m)=C_{4m-4}(m)= \frac{3 (-1)^m m (2 m (2 m (2 m (8 m (6 m-7)+155)-13)-1017)-315)}{19600
   \left(m-\frac{3}{2}\right) \left(m-\frac{1}{2}\right)
   \left(m+\frac{1}{2}\right) \left(m+\frac{3}{2}\right)}.
\end{equation}
These results were obtained using the  "rate", guessing program of C. Krattenthaler. Then further, M. Trott was able to obtain the result 
(using the FindSequenceFunction command of Mathematica, which searches for a possible rational form)
\begin{equation} \label{coefficient6}
C_{6}(m)=C_{4m-6}(m)=
\end{equation}
\begin{displaymath}
\frac{(-1)^m (m-1) m (4 m (2 m (2 m (m (4 m (20 m (4
   m-11)+173)-4303)+4733)+14911)-9165)-4725)}{529200
   \left(m-\frac{5}{2}\right) \left(m-\frac{3}{2}\right)
   \left(m-\frac{1}{2}\right) \left(m+\frac{1}{2}\right)
   \left(m+\frac{3}{2}\right)}.
\end{displaymath} 

From the formulas for these coefficients, it is 
clear that the numerator of the coefficient $C_{2 j}(m) (=C_{4 m-2j}(m)$) 
of $\mu^{2 j}$ is
a polynomial of degree  $3 j$, and the denominator
is a polynomial of degree $j+2$. (The denominators are very simple in structure (\ref{hammer})--as evidenced above.). For $j=0$, the roots are $-\frac{3}{2}$ and 
$-\frac{1}{2}$, and as $j$ increases by 1, an additional root 1 larger in value than the previous smallest is added. Thus, {\it poles} occur at the coefficient functions at such half-integers.) 
\subsubsection{Asymptotic convergence of dominant roots to half-integrs}
Utilizing this observation, we were then able to move on to obtaining the coefficients 
$C_{8}(m)=C_{4m-8}(m)$,  
$C_{10}(m)=C_{4m-10}(m)$ $C_{12}(m)=C_{4m-12}(m)$,  
$C_{14}(m)=C_{4m-14}(m)$ and 
$C_{16}(m)=C_{4m-16}(m)$--but not yet higher. In studying the (nontrivial) roots of these functions, we have detected one quite interesting feature. That is, as $j$ increases, the dominant roots of $C_{2j}(m)$ show very strong evidence of converging 
to $j-\frac{1}{2}$, the subdominant roots 
to $j-\frac{3}{2}$,...For instance, for $j=8$, the dominant roots of $C_{16}(m)=
C_{4m -16}(m)$ are $7.49999796, 6.4999352, 5.4980028, 4.4493216$, while for $j=7$, they are $6.5000204$, $5.500556$, $4.515944$. Such roots would then come increasingly close to canceling the near-to-matching poles in the denominators in $C_{2 j}(m)$ as $j$ increases. (This behavior suggests an intimate connection with the theory of angular momentum and its 
[semiclassical] asymptotics \cite{Exact}.)
\subsection{Use of the intermediate functions $I_{m}(\mu)$ to compute the $m$-th moment of $|\rho^{PT}|$}
In the second stage of our procedure to compute the $m$-th Hilbert-Schmidt moment, we reverse the substitution (\ref{substitution}) 
in these $4 m$-degree polynomials,
multiply the result by the necessarily {\it nonnegative} factor 
$(\rho_{22} \rho_{33})^{2 m}$ (the factor $(\rho_{22} \rho_{33})^2$ had been removed, as previously indicated, in forming the polynomial $P$ in seven variables, proportional to $|\rho^{PT}|$) and also by the jacobian corresponding to the transformation to Bloore (correlation) variables for the two-rebit density matrices \cite{andai}
\begin{equation} \label{andaiDirichlet}
jac= (\rho_{11} \rho_{22} \rho_{33} \rho_{44})^{\frac{\beta}{2}}, 
\hspace{.2in} \beta=3.
\end{equation}
($\beta=6$, for the two-qubit case, and $\beta=12$ for the corresponding quaternionic scenario, in accordance with random matrix theory results.)
The result of the indicated multiplications is, then, integrated over the unit three-dimensional simplex,
\begin{equation}
\rho_{11}+\rho_{22}+\rho_{33} +\rho_{44}=1, \hspace{.2in} \rho_{ii} \geq 0, \hspace{.2in} i=1,\ldots,4
\end{equation}
to obtain the $m$-th moment ($\zeta_m^{'}$). In other words (taking into account the appropriate normalization factor), and setting 
$\rho_{44}=1-\rho_{11}-\rho_{22}-\rho_{33}$, the computation takes the form
\begin{equation} \label{normfactor}
Moment_{m} \equiv \zeta_m^{'}=\frac{1146880}{\pi ^2} \int_0^1 \int_0^{1-\rho_{11}} \int_0^{1-\rho_{11}-\rho_{22}} 
\end{equation}
\begin{displaymath}
(\rho_{22} \rho_{33})^{2 m} (\rho_{11} \rho_{22} \rho_{33} \rho_{44})^{\frac{3}{2}} I_{m}(\sqrt{\frac{\rho_{11} \rho_{44}}{\rho_{22} \rho_{33}}}) \mbox{d} \rho_{33} \mbox{d} \rho_{22} 
\mbox{d} \rho_{11}. 
\end{displaymath}
We are, in fact, able to  perform the indicated {\it symbolic} integration, obtaining 
thereby (using an index $i \equiv 2 j$)
\begin{equation} \label{MomentForm1}
Moment_{m}=\zeta_m^{'}=\frac{1146880}{\pi ^2 \Gamma (4 m+10)} \Sigma_{i=0,2,4...}^{4 m} 
\Gamma \left(\frac{i+5}{2}\right)^2 \Gamma \left(-\frac{i}{2}+2
   m+\frac{5}{2}\right)^2 C_{i}(m)
\end{equation}
\begin{displaymath}
=\frac{2293760}{\pi ^2 \Gamma (4 m+10)} \Big( \Sigma_{i=0,2,4...}^{2 m-2} 
\Gamma \left(\frac{i+5}{2}\right)^2 \Gamma \left(-\frac{i}{2}+2
   m+\frac{5}{2}\right)^2 C_{i}(m)\Big) +
\end{displaymath}
\begin{displaymath}
+\frac{1146880}{\pi ^2 \Gamma (4 m+10)} \Gamma \left(m+\frac{5}{2}\right)^4 C_{2 m}(m),
\end{displaymath}
where the  $C_{i}(m) (\equiv C_{2j}(m))$'s  are our previously-indicated rational coefficient functions 
((\ref{coefficient0})-(\ref{coefficient6})), symmetric about
$2 m$. These (rational functions) $C_{i}(m)$'s 
themselves--as discussed above--are the ratios of polynomials in $m$ of degree $\frac{3 i}{2}$ divided by 
 the term (using the Pochhammer symbol, as well as rising factorials for gamma functions with half-integer arguments)
\begin{equation} \label{hammer}
\mbox{denominator}(C_{i}(m)) 
=\left(\frac{1-i}{2}+m\right)_{\frac{i}{2}+2} = 
\frac{\Gamma \left(m+\frac{5}{2}\right)}{\Gamma
   \left(-\frac{i}{2}+m+\frac{1}{2}\right)}=
\frac{2^{\frac{i}{2}+2} (2 m+3)\mbox{!!}}{(-i+2 m-1)\mbox{!!}}
\end{equation}
\begin{displaymath}
=\Pi_{k=-2,0,...}^{i} (m+\frac{1-k}{2}).
\end{displaymath}
For $i=4$, by way of example, this gives us the denominator of (\ref{coefficient4}),
\begin{equation}
\left(m-\frac{3}{2}\right) \left(m-\frac{1}{2}\right)
   \left(m+\frac{1}{2}\right) \left(m+\frac{3}{2}\right).
\end{equation}
On the other hand, the {\it numerators} of the $C_{i}(m)$'s for $m>0$
have zero as a trivial root, and for $m> 4 n$, 
trivial roots $0, \ldots n$.

Again, converting gamma functions with half-integer arguments to rising factorials, we have, equivalently to (\ref{MomentForm1}), that 
\begin{equation}
Moment_m=\zeta_m^{'}=
\end{equation}
\begin{displaymath}
35 \frac{ 2^{7-4m}}{ \Gamma[4 m+10]} \Big[ \Big((2m+3)!!\Big)^2 C_{2 m}(m) +2 \Sigma_{i=0,2,4...}^{2 m-2}
\Big((3+i)!! (3-i+4 m)!!\Big)^2 C_i(m) \Big].
\end{displaymath}

Pursuant to these formulas, the first moment (mean) of the Hilbert-Schmidt probability distribution of $|\rho^{PT}|$
over the interval $[-\frac{1}{16},\frac{1}{256}]$  was found to be
\begin{equation} \label{firstmoment}
\zeta_1^{'}=-\frac{1}{858} =-\frac{1}{2 \cdot 3 \cdot 11 \cdot 13} 
\approx -0.0011655,
\end{equation}
falling within the [negative] region ($|\rho^{PT}|<0$) 
of entanglement. (We were also able to compute this result using the alternative Euler-angle parameterization of the real density matrices \cite{JMP2008}, but only after correcting a typographical error in the associated Haar measure \cite[eq. (48)]{JMP2008}, in which the factor $\sin{x_3}$ had to be replaced by its square. Also, we depart from the standard convention of denoting moments by
$\mu$, since that symbol has been employed in our earlier studies \cite{slaterPRA2,advances} 
and above (\ref{substitution}).) Then, successively,  the ([necessarily] decreasing in absolute value) raw (non-central) moments are
\begin{equation} \label{MomentTwo}
\zeta_2^{'}=\frac{27}{2489344} =\frac{3^3}{2^{10} \cdot 11 \cdot 13 \cdot 17}\approx 0.0000108462,
\end{equation}
\begin{equation}
\zeta_3^{'}=-\frac{8363}{66216550400} = -\frac{8363}{2^{13} \cdot 5^2 \cdot 7 \cdot 11 \cdot 13 \cdot 17 \cdot 19} \approx -1.2629773 \cdot 10^{-7},
\end{equation}
\begin{equation}
\zeta_4^{'}=\frac{21859}{10443295948800} = 
\frac{21859}{2^{17} \cdot 3 \cdot 5^{2} \cdot 11 \cdot 13 \cdot 17 \cdot 19 \cdot 23} \approx 2.09311 \cdot 10^{-9},
\end{equation}
\begin{equation}
\zeta_5^{'}= -\frac{23071}{539633583390720} =- \frac{23071}{2^{18} \cdot 3 \cdot 5 \cdot 7^2 \cdot 13 \cdot 17 \cdot 19 \cdot 23 \cdot 29}\approx-4.27531 \cdot 10^{-11},
\end{equation}
\begin{equation}
\zeta_6^{'}=\frac{3317321}{3253917653076541440} = 
\frac{7 \cdot 43 \cdot 103 \cdot 107}{2^{28} \cdot 3 \cdot 5 \cdot 11^2 \cdot 17 \cdot 19 \cdot 23 \cdot 29 \cdot 31} \approx 1.01949 
\cdot 10^{-12},
\end{equation}
\begin{equation}
\zeta_7^{'}=-\frac{419856257}{15366774022001834065920} =
\end{equation}
\begin{displaymath}
-\frac{43 \cdot 2179 \cdot 4481}{2^{30} \cdot 3^4 \cdot 5 \cdot 11 \cdot 13 \cdot 17 \cdot 19 \cdot 23 \cdot 29 \cdot 31 \cdot 37}
\approx -2.73223 \cdot 10^{-14},
\end{displaymath}
\begin{equation} 
\zeta_8^{'}=\frac{16945249}{21117403549591928832000} =
\end{equation}
\begin{displaymath}
\frac{109 \cdot 155461}{2^{33} \cdot 3 \cdot 5^3 \cdot 11  \cdot 19 \cdot 23 \cdot 29 \cdot 31 \cdot 37 \cdot 41}
\approx 8.02431  \cdot 10^{-16},
\end{displaymath}
and (requiring four days of Mathematica computation on a MacMini machine)
\begin{equation} \label{lastmoment}
\zeta_9^{'}=-\frac{6102620963}{240565904621616585139814400} =
\end{equation}
\begin{displaymath}
-\frac{19 \cdot 199 \cdot 1614023}{2^{37} \cdot 3 \cdot 5^2 \cdot 11^3  \cdot 13  \cdot 23 \cdot 29 \cdot 31 \cdot 37 \cdot 41 \cdot 43}
\approx -2.53678  \cdot 10^{-17}.
\end{displaymath}
(After four weeks of uninterrupted computation, we did not succeed, however, in determining $\zeta_{10}^{'}$.)

Interestingly, the sequence of denominators  immediately above 
(in apparent contrast to that of the numerators) appears to 
be "nice" in that the number of their prime factors do not grow rapidly, but rather linearly. This is a strong indication of the possible existence of 
a "closed form", that is 
an expression which is built by forming products and quotients of 
factorials \cite[fn. 12]{determinantcalculus}.

\subsection{Use of moments to estimate the  probability distribution over $|\rho^{PT}|$}
\begin{figure}
\includegraphics[scale=2]{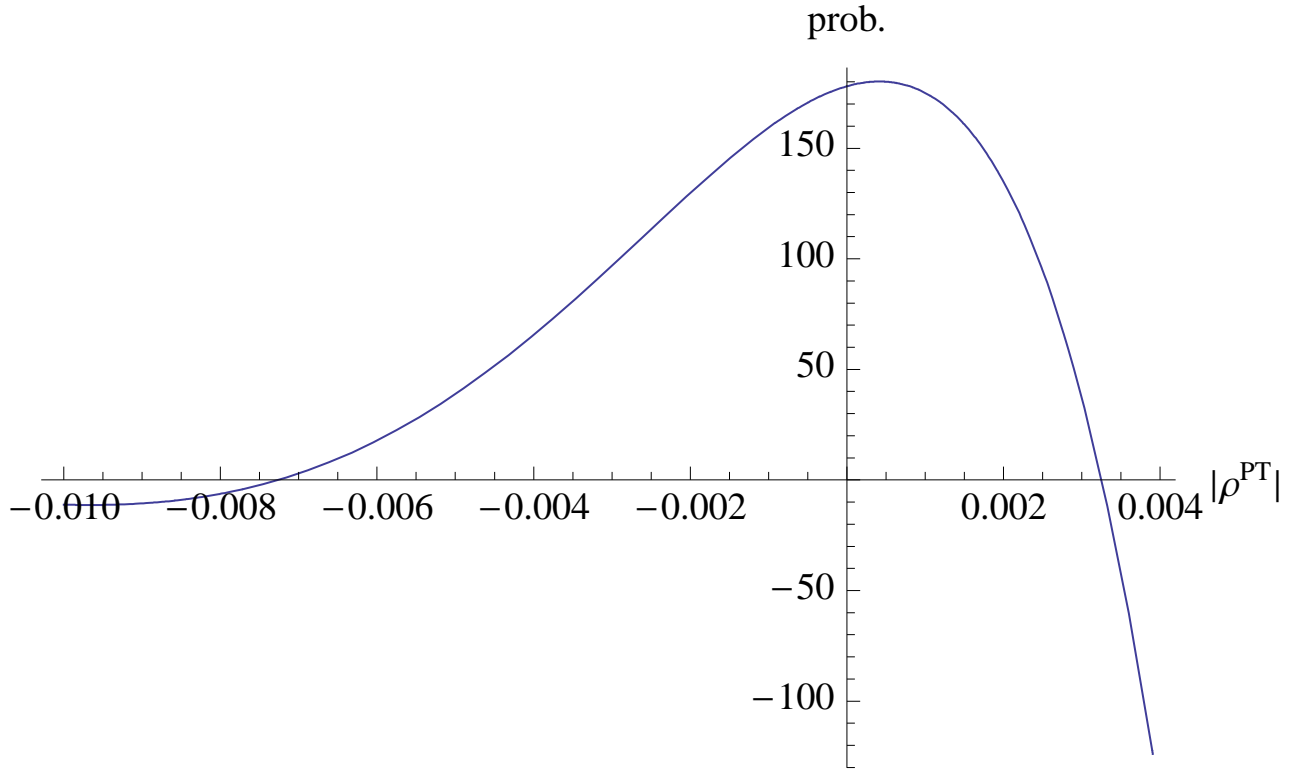}
\caption{\label{fig:TransposeFit}Fit--without nonnegativity constraints imposed--of a nine-degree polynomial to the first nine exactly-computed moments of the Hilbert-Schmidt probability distribution over $|\rho^{PT}|$, where $\rho$ is a generic two-rebit density matrix.  The domain of separability is 
$|\rho^{PT}| >0$.}
\end{figure}

In Fig.~\ref{fig:TransposeFit}, we display a (naive) fit of a 
simple power series in 
$|\rho|^{PT}$ of degree nine to the computed first nine moments 
((\ref{firstmoment})-(\ref{lastmoment})) (cf. \cite[Figs. 1, 2]{giraud1}) 
of the Hilbert-Schmidt probability distribution over $|\rho^{PT}|$, where  
$\rho$ is a generic two-rebit density matrix.
No nonnegativity constraints were, however, imposed and considerable
incursions into negative regions result. (Such negativity can be obviated through the use of maximum-entropy, spline-fitting and other methodologies 
\cite{john,mnatsakanov2,mnatsakanov,bertsimas,popescu,HP}, and we do explore such directions.)

The Hilbert-Schmidt separability probability predicted by the curve in 
Fig.~\ref{fig:TransposeFit}--that is the "probability mass" (the resultant of both positive and negative values) lying within the interval $[0,\frac{1}{256}]$--is $0.39648$, while our previous studies 
\cite{advances}, indicate that the actual value is somewhat higher, $\approx 0.45$--a discrepancy the use of additional higher-order moments should ameliorate.
 
Since the plotted distribution (Fig.\ref{fig:TransposeFit}) appears to be {\it unimodal}, one can presumably use the computations of the first and second moments above to isolate the mode of the distribution within the interval \cite[eq. (13)]{gavriliadis1}
\begin{equation}
\left\{-\frac{1}{858}-\frac{\sqrt{\frac{30397}{51}}}{4576},
\frac{\sqrt{\frac{30397}{51}}}{4576}-\frac{1}{858}\right\} =
\{-0.00650062, 0.00416962\},
\end{equation}
containing the value $|\rho^{PT}|=0$. Narrower intervals containing the mode can be obtained using
higher-order moments and the associated Hankel determinants 
\cite[Thm. 3.2]{gavriliadis1}.
\subsection{Numerical computations of higher-order intermediate functions ($j \geq 9$)}
It is clear that it would be of considerable utility  to have available
exact values for still higher-order (than $m=9$) moments--and for the coefficients $C_{2j}(m) 
\equiv C_{i}(m)$ of the terms in the intermediate functions/polynomials  
$I_{m}(\mu)$--but the associated computational demands seem quite considerable. 

Our only current 
recourse, in this regard, then, appeared to be a numerical one.
We employed a quasi-Monte Carlo (Tezuka-Faure \cite{giray1,tezuka}) procedure
(using 18,870,000 [low-discrepancy] points) to estimate the values of $C_{2j}(m)$ for $j\geq 9$ and $m \geq 10$, 
for $m=10,\ldots 50$. (Actually, by the evident symmetry around the $2m$-th power of the coefficients of the intermediate functions, we were able to obtain two values [which we averaged] for each point.) Coupling these approximate results with the exact formulas for the intermediate functions obtained above for $j<9$ (sec.~\ref{InterSection}), we obtained estimates of the first {\it fifty} moments. (We investigated
the possibility of using these additional numerical results to infer
the desired further {\it exact} formulas for the $C_{2j}(m)$'s, but this seemed too demanding a task, given the ordinary machine precision of our quasi-Monte Carlo simulations and the evident high [multi-digit] complexity of the exact coefficients.)
\subsubsection{Cumulative distribution function calculations}
We then used these values (the first nine moments being exact, and the remaining "semi-exact" forty-one, being 
sums of exact and numerical terms) in the procedure of
Mnatsakanov for approximating the "moment-determinate cumulative distribution function (cdf) from its moments". 
The relevant formula for the cdf (at the separable/nonseparable boundary 
of principal interest to us) based on the first $K=m$ moments (linearly mapping $|\rho^{PT}|$ to lie 
in [0,1], and the moments accordingly) is of the form  \cite[eq. (2)]{mnatsakanov2} \cite[eq. (1.3)]{gzyl}\begin{equation}
F_{K,\zeta'}
= \Sigma_{k=0}^{\lfloor {\frac{16 K}{17}} \rfloor}  \Sigma_{j=k}^K (-1)^{j-k} \binom{K}{j} \binom{j}{k}  \zeta_j^{'}  .
\end{equation}
(We have to subtract this [entangled probability estimate] from 1 to get the [complementary] separability probability estimate, to be plotted as a function of the number of moments $K$.) 

\begin{figure}
\includegraphics[scale=2]{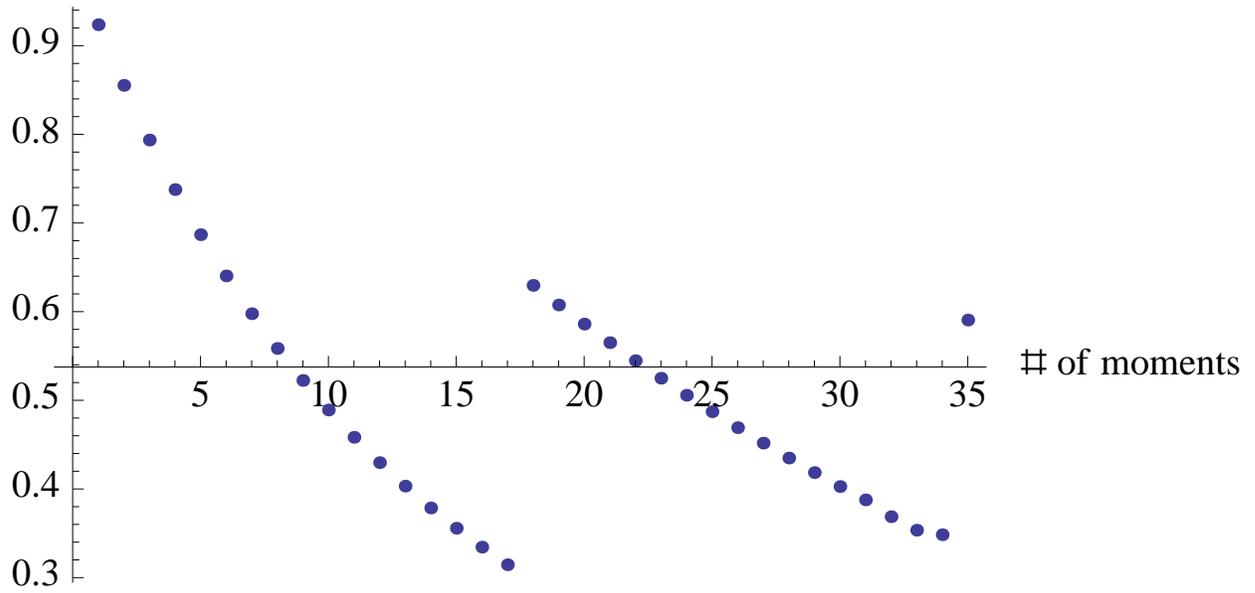}
\caption{\label{fig:CDFmnats}Separability probability estimates based on
differing numbers (exact for $m \leq 9$ and "semi-exact" for $m>9$) of moments, using the reconstruction procedure of Mnatsakanov \cite{mnatsakanov2}. The horizontal axis is drawn to intercept the vertical at  
$\frac{1129}{2100} \approx 0.537619$, the least upper bound so far established \cite{advances}.}
\end{figure}

\begin{figure}
\includegraphics[scale=2]{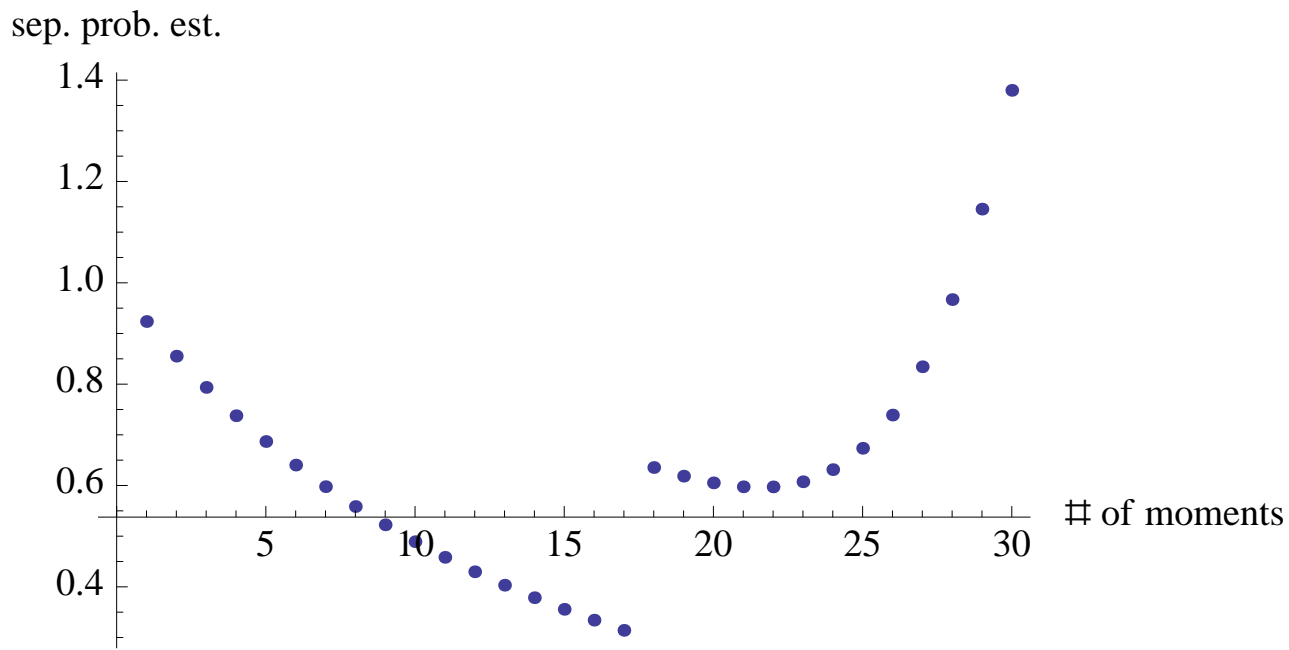}
\caption{\label{fig:CDFmnats2}Separability probability estimates, without any complementary numerical (quasi-MonteCarlo) input (for $m>9$), using the reconstruction procedure of Mnatsakanov \cite{mnatsakanov2}.  The horizontal axis intercepts the vertical at  $\frac{1129}{2100} \approx 0.537619$, the least upper bound so far established \cite{advances}.}
\end{figure}

In Fig.~~\ref{fig:CDFmnats}, we show estimates of this value based on increasingly large numbers of moments. (For more than thirty-six moments, the results turn negative. If we just employ in this same reconstruction procedure, the known exact formulas for the coefficients of the intermediate functions--effectively setting the supplementary/corrective numerical terms to zero--then the two sets of results are quite close up to twenty moments, but then become considerably more ill-behaved when none of our supplementary numerical results is included [Fig.~\ref{fig:CDFmnats2}]. In a recent interesting study, Gzyl and Tagliani concluded that thirty-two was "the maximum allowable moments before incurring numerical instability, unless one conducts the calculation with high accuracy \cite{gzyl}. They also assert that "the additional information introduced by using the $(M+1)$-order moment is 'visible' only after the $0.6M$-th decimal digit". In this regard, let us note that the quasi-Monte Carlo calculations used to complement the exact results here were conducted with only ordinary machine precision.)

\begin{figure}
\includegraphics[scale=2]{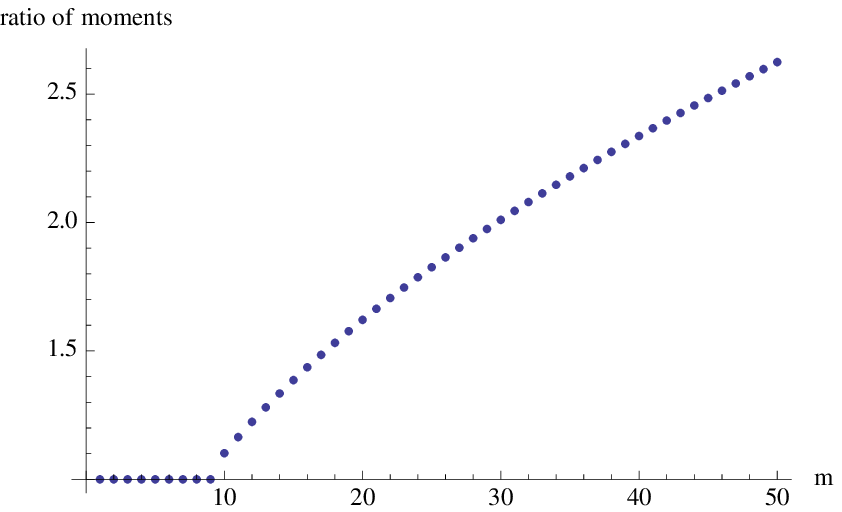}
\caption{\label{fig:MomentRatios}Ratios of moments computed using the exact and complementary numerical (for $m>9$) results to those based only on the exact, but incomplete (for $m>9$) results}
\end{figure}

It would appear that the wide range  and lack of stability of estimates 
is reflective
of the ill-posedness of the Hausdorff moment (inverse) problem (stemming ultimately
from the lack of orthogonality of the sequence $1, x,\ldots,x^n,\ldots$ 
\cite{gzyl}). In Fig.~\ref{fig:MomentRatios} we show the ratio of the exact (but incomplete for $m>9$) moment computations to that (semi-exact one) based on the exact and complementary numerical results.
\subsection{Libby-Novick (three-parameter) probability distribution}
We have explored the use of extensions of the (two-parameter) beta distribution 
\cite[chap. 5]{handbook} \cite{gordy} to better fit the nine exact moments than found, as presented above (sec.~\ref{BetaSection}), with the particular distribution (\ref{goodbetafit2})
that had been fit to the first two moments of $|\rho^{PT}|$.
Doing so, we were able to obtain a {\it three}-parameter Libby-Novick (LN) distribution of 
the form \cite{libby} \cite[eq. (IX.1)]{handbook},
\begin{equation} \label{LibbyNovick}
P_{LN}(y)=\frac{\lambda^{a} y^{a-1} (1-y)^{b-1}}{B(a,b)(1-(1-\lambda)y)^{a+b}},
\end{equation}
\begin{displaymath}
a \approx 3.7141606, b \approx 359.577737, 
\lambda \approx 0.00064805.
\end{displaymath}
This gave us a further considerably improved fit over that of (\ref{goodbetafit2})
to the first nine exact moments.
(The ninth moment was now predicted within $99.6\%$--as opposed to within $98.6\%$--and the preceding moments
better still--in a monotonically declining goodness-of-fit manner from the first to the ninth.)
The estimated separability probability now increased to 0.429121.

To arrive at the probability distribution (\ref{LibbyNovick}), we started with the beta distribution fit (\ref{goodbetafit2}) that exactly reproduced
the first two moments, now trying to fit the first {\it three} moments. 
This involved a very long (slowly converging) 
iterative process, which at each stage, 
appealingly, seemed to improve the fit to all our computed fifty
(nine exact and forty-one, "semi-exact") moments. Additionally, the estimated separability probability seemed to increase at each step, 
which we found to better accord with our earlier extensive numerical investigations \cite[sec. IX.A]{slater833} 
\cite[sec. V.A.2]{slaterPRA2} \cite{advances}.

One can employ the Libby-Novick distribution (\ref{LibbyNovick}) as 
a "baseline density", in the manner below using the beta distribution, 
following the methodology of Ha and Provost 
\cite[eq. (3.5)]{HP}, to generate estimates of the HS 
separability probability.
\subsection{Moments based on lesser principal minors of $\rho^{PT}$ than its determinant}
A necessary, but not sufficient condition that a two-qubit density matrix 
be separable is that any $3 \times 3$ principal minor of its partial transpose be nonnegative \cite{advances}. So, we can select one such minor, say 
(cf. (\ref{PT})),
\begin{equation}
\mbox{minor}_{3 \times 3} = 
\frac{\rho_{11}^2 \rho_{44}}{\mu^2} (\mu^2 z_{14}^2 -2 
\mu z_{12} z_{13} z_{14} +z_{13}^2 +z_{12}^2 -1),
\end{equation}
expressed in terms of the Bloore parameterization, and compute the associated moments. (Enforcing the nonnegativity of such a minor yields an {\it upper bound} of $\frac{22}{35} \approx 0.628571$ on 
the Hilbert-Schmidt probability of separability of generic two-rebit systems \cite{advances}.) We have been able to compute exactly the first forty-five such moments of the probability distribution over the interval 
$[-\frac{1}{8},\frac{1}{27}]$, following very much the same scheme
as we pursued  for the first nine moments of $|\rho^{PT}|$. The first two moments are $-\frac{1}{264}$ and $\frac{7}{74880}$, while remarkably, the (raw) third moment is identically zero (cf. \cite{elkin}). The form the corresponding "intermediate functions" now took for these first three cases 
were (cf. (\ref{inter1})-(\ref{inter3}))
\begin{equation}
I_{1}(\mu)=\frac{1}{5} \left(\mu ^2-3\right),
\end{equation}
\begin{displaymath}
I_{2}(\mu)=\frac{1}{875} \left(75 \mu ^4-182 \mu ^2+395\right)
\end{displaymath}
and
\begin{displaymath}
I_{3}(\mu)=\frac{125 \mu ^6-297 \mu ^4+675 \mu ^2-935}{2625}.
\end{displaymath}
General formulas could now also be obtained for the coefficients of the $2j$-th powers of the intermediate functions. These involved the Mathematica functions  LerchPhi or DifferenceRoot (cf. (\ref{coefficient0})-(\ref{coefficient6})). 
However, we were unable to find 
a general formula for the $m$-th moment, even based on the availability of the first fifty-eight moments. 

\section{Moments of the product of $|\rho|^{PT}$ and $|\rho|$} 
\label{ProductSection}
The main foci of our initial analyses had been, firstly, the moments of the determinant $|\rho^{PT}|$ of the 
partial transpose of generic two-rebit density matrices, and, secondarily, the moments of
the (necessarily nonnegative) determinant $|\rho|$ 
of the underlying density matrix itself--the moments of 
the latter determinant being more 
computationally amenable, it is clear, to exact analyses.

Now, in this context, it is not unnatural to ask the question of the nature of the
moments of the {\it product} of these two determinants, 
 $|\rho| |\rho^{PT}|=|\rho \rho^{PT}|$ (cf. \cite[p. 564]{fyodorov}). This approach potentially contributes
insight into the separability probability question, since a value of the product less than zero still
indicates the presence of an entangled state, and a value greater than zero, a separable state. (It would appear that the argument in \cite{augusiak} of Augusiak, Horodecki and Demianowicz
could be adapted, so that the value of the product could be construed as that obtainable from a single certain observable measurement.)

In undertaking the associated analysis of moments, we immediately encountered a most interesting result. The first moment
or mean of the normalized product $|\rho| |\rho^{PT}|$ is zero, that is
\begin{equation} \label{ZeroMean}
\zeta_{1}^{'}=0,
\end{equation}
the associated "intermediate function" being (cf. (\ref{inter1}))
\begin{equation} \label{InterOne}
I_{1}(\mu)=-\frac{24 \mu ^4}{875}+\frac{3888 \mu ^2}{42875}-\frac{24}{875}.
\end{equation}
(Numerically, the first moment of the {\it absolute} value of the normalized product is $5.86519 \cdot 10^{-7}$. This is close to the absolute value, $\frac{1}{1963104} \approx 5.09397 \cdot 10^{-7}$, of the product of the mean of $|\rho|$, that is, $\frac{1}{2288}$ and the mean of $|\rho^{PT}|$, $-\frac{1}{858}$.)
The second moment (cf. (\ref{MomentTwo})) is
\begin{equation}
\zeta_{2}^{'}=\frac{7}{5696343244800}=\frac{7}{2^{18} \cdot 3^2  \cdot 5^2 \cdot 13 \cdot 17 \cdot 19 \cdot 23} \approx 1.2288585 \cdot 10^{-12}
\end{equation}
(so, the corresponding 
standard deviation is the square root of this, that is 
$\frac{\sqrt{\frac{7}{96577}}}{7680} \approx 1.10854 \cdot 10^{-6}$) with the associated intermediate function being (cf. (\ref{inter2}))
\begin{equation} \label{InterTwo}
I_{2}(\mu)=\frac{192 \mu ^8}{94325}-\frac{12032 \mu ^6}{1528065}+\frac{5561984 \mu
   ^4}{184895865}-\frac{12032 \mu ^2}{1528065}+\frac{192}{94325}.
\end{equation}
(To compute the $m$-th moment of the probability distribution of the product, 
using the new set of intermediate functions, we employ the same formula as (\ref{normfactor}), 
but for the replacement of the exponent $\frac{3}{2}$  
by $\frac{3}{2}+m$.) We see that the coefficients of the constant (and highest power of $\mu$)
terms in the first and second intermediate functions immediately above 
((\ref{InterOne}), (\ref{InterTwo})) are 
$-\frac{24}{875}$ and $\frac{192}{94325}$. The next four such 
constant coefficients ($m=3, 4, 5, 6$) have been found to 
be $-\frac{1024}{4729725}, \frac{16384}{586831245}, -\frac{393216}{96770250577}$ and 
$\frac{1048576}{1631366611875}$. However, we have yet to obtain a general formula, parallel 
to (\ref{coefficient0}), in light of the increased computational burden, encompassing these six values.

The third to the sixth moments  are
\begin{equation}
\zeta_{3}^{'}=\frac{1}{677899511057612800}=\frac{7}{2^{18} \cdot 3^2  \cdot 5^2 \cdot 13 \cdot 17 \cdot 19 \cdot 23} \approx 1.2288585 \cdot 10^{-12}
\end{equation}
\begin{equation}
\zeta_{4}^{'}=\frac{1}{45973294808920227840000}=\frac{7}{2^{18} \cdot 3^2  \cdot 5^2 \cdot 13 \cdot 17 \cdot 19 \cdot 23} \approx 1.2288585 \cdot 10^{-12}
\end{equation}
\begin{equation}
\zeta_{5}^{'}=\frac{1}{11662680803407302839532257280}=\frac{7}{2^{18} \cdot 3^2  \cdot 5^2 \cdot 13 \cdot 17 \cdot 19 \cdot 23} \approx 1.2288585 \cdot 10^{-12}
\end{equation}
and
\begin{equation}
\zeta_{6}^{'}=\frac{3929}{4158654163938276392103553381781471232}=\frac{7}{2^{18} \cdot 3^2  \cdot 5^2 \cdot 13 \cdot 17 \cdot 19 \cdot 23} \approx 1.2288585 \cdot 10^{-12}.
\end{equation}

Despite their lengthy digital descriptions, the ratios of these six moments
to the HS moments of $|\rho|^{2 k}$--given by (\ref{generalformula}) are rather remarkably simple, that is,
\begin{equation}
\{ 0, \frac{77}{54}, \frac{24}{55}, \frac{209}{175}, \frac{598}{833}, \frac{3929}{3724} \}
\end{equation}
We have only so far been able to compute the two-qubit analogue of the first (zero) of the six ratios above--it turning out quite remarkably 
to be $\frac{3}{2}$/
Since these ratios are so simple, it suggested to us that we might be more able to progress in our series of analyses, by making our initial goal
the computation of  these (unknown) but apparently well-behaved 
ratios--for higher-order moments--rather than the very small values of these moments themselves.

\section{Summary}
We have studied here the moments of probability distributions generated by
certain determinantal functions of generic two-qubit density matrices 
($\rho$) with real
entries ("rebits") over the associated nine-dimensional convex domain, assigned  Hilbert-Schmidt measure. It was found 
that the mean of the (nonnegative) determinant $|\rho|$ is 
$\frac{1}{2288}$, the mean of the determinant 
 of the partial transpose $|\rho^{PT}|$--negative values 
indicating entanglement--is
$-\frac{1}{858}$, while the mean of the product of these two determinants,  $|\rho| |\rho^{PT}|=|\rho \rho^{PT}|$, is
{\it zero}. We  determined the exact values--also rational numbers--of the succeeding eight moments of 
$|\rho^{PT}|$.  At intermediate steps in the derivation of the $m$-th moment of $|\rho^{PT}|$, rational functions 
$C_{2 j}(m)$ emerge, yielding the coefficients of the $2j$-th  
power of even polynomials ("intermediate functions" $I_{m}(\mu)$) 
of total degree $4 m$. These functions  possess poles at finite series of consecutive half-integers 
($m=-\frac{3}{2},-\frac{1}{2},\ldots,\frac{2 j-1}{2}$), and certain 
(trivial) roots at finite series of consecutive natural numbers ($m=0,\ldots, \left\lfloor \frac{m}{4}\right\rfloor$). 
The (nontrivial)
dominant roots of  $C_{2 j}(m)$ appear to converge, as $j$ increases, to 
the same half-integer values 
($m=\ldots,\frac{2 j-3}{2}, \frac{2 j-1}{2},$). If formulas for $C_{2j}(m)$ could be developed for arbitrary $j$--we do possess them 
already for $j<9$--then, the 
desired Hilbert-Schmidt separability probability would be computable to high accuracy.

We reproduced the (linearly transformed) first nine moments of $|\rho^{PT}|$ quite  
closely by a certain (two-parameter) beta distribution, and still more closely by a three-parameter (Libby-Novick) extension of it. The first two moments of $|\rho^{PT}|$--when employed in the 
one-sided Chebyshev inequality--gave an upper bound
of $\frac{30397}{34749} \approx 0.874759$ on the Hilbert-Schmidt separability probability 
 of two-rebit density matrices. 
We ascertained by numerical  methods that the orthogonality established of 
$|\rho|$ and $|\rho^{PT}|$ with respect to Hilbert-Schmidt  measure does not hold with respect to the Bures (minimal monotone) measure, nor if we 
slightly distort the symmetry of our basic nine-dimensional generic two-rebit scenario.


\begin{acknowledgments}
I would like to express appreciation to the Kavli Institute for Theoretical
Physics (KITP)
for computational support in this research, to Michael Trott
for lending his Mathematica expertise, and Christian Krattenthaler, Mihai Putinar, Robert Mnatsakanov, Mark Coffey and Charles Dunkl for general discussions and insights.
\end{acknowledgments}

\end{document}